# The relationship between the US broad money supply and US GDP for the time period 2001 to 2019 with that of the corresponding time series for US national property and stock market indices, using an information entropy methodology

Laurence Francis Lacey

Lacey Solutions Ltd, Skerries, County Dublin, Ireland

June 03 2021

# The relationship between the US broad money supply and US GDP for the time period 2001 to 2019 with that of the corresponding time series for US national property and stock market indices, using an information entropy methodology

## Abstract

The primary objective of this paper was to investigate whether the growth in the major US asset indices could be a function of the US broad money supply and/or US GDP, over the time period 2001 to 2019, using an information entropy methodology. The four US asset indices investigated were: (1) US National Property index; (2) Russell 2000 index; (3) S&P 500 index; and (4) NASDAQ index. Notwithstanding the financial crisis of 2007-2008, US real GDP increased exponentially over the period 2001 to 2019, with an average annual growth rate of approximately 2%. However, over this time period, the average annual rate of growth of US GDP was considerably lower than the average annual rate of growth of the US broad money supply (5.7%). The main determinant of the average growth rate for all four US asset indices studied would appear to be the growth rate in the US broad money supply. In addition, the growth rate in the US Russell 2000 stock index and the NASDAQ index would appear to be a function of the combined positive effects of both the growth rate in the US Broad Money Supply and the growth rate of US GDP.

## 1. Introduction

A new information entropy (Info Ent) methodology [1] was applied to investigate the relationship between the US broad money supply and the US GDP for the time period 2001 to 2019, with 2001 as the reference year (time = 0), with the following 2001 to 2019 time series for the indices measuring the increase in:

(1) Average annual US national property prices

(2) Russell 2000 (smaller companies) annual stock prices

(3) S&P 500 (large companies) annual stock prices

(4) NASAQ (large technology) annual stock prices

## 2. Methods

An information entropy statistical methodology has been developed for investigating expansionary processes, in with full details for an exponential expansionary process can be found [1]. For an exponential expansionary process, the methodology provides a rate-constant (λ) for the exponential growth of the process (G(t)) and the associated information entropy for the time series under investigation [1]. This can be expressed as follows:

$$\mathrm{G(t)} = \exp(\lambda \, \mathrm{x} \, \mathrm{t})$$

and,

$$\text{Info Ent } (\mathrm{G(t)}) = \lambda \, \mathrm{x} \, \mathrm{t}$$

While information entropy has no units, at any given time, it is related to the average growth rate ($r$), where:

$$r = \exp(\lambda) - 1 \quad \text{and} \quad \lambda = log_e(1 + r)$$

Consequently, information entropy can be considered to be related to the "velocity" of the growth of time series.

The following *a priori* hypotheses were examined:

(1) The increase in the average annual US property prices is a function of the increase in US GDP. More specifically:

$$Info\ Ent\ (US\ property\ prices) = Info\ Ent\ (US\ GDP)\ \pm constant$$

(2) The increase in the annual US stock market indices is a function of the increase in US GDP and the US money supply. More specifically:

$$Info\ Ent\ (US\ stock\ market\ index) = Info\ Ent\ (US\ Broad\ Money\ Supply)\ \pm Info\ Ent\ (US\ GDP) \pm constant$$

All results of the data analysis given below were obtained using Microsoft Excel 2019, 32-bit version.

# 3. Results

## 3.1 Information Entropy of the US broad money supply (2001 to 2019)

The broad money supply of US$ from 2001 to 2019 (time, 0 to 18 years) [2] is plotted in Figure 1. As can been seen, the expansion of the money supply can be well described using an exponential function, in which,

$$US\ Broad\ Money\ supply = 7.5805\ trillion\ US\$, when\ t = 0\ (2001)$$

The coefficient of determination ($R^2$) of the fit is approx. 98% for the 19-year time series. Over this period, the exponential rate-constant (λ) = 0.0555. This is equivalent to a mean annual rate of increase in the broad money supply of US$ over this time period of 5.7%.

Figure 1: Mono-exponential expansion of the broad money supply of US$ from 2001 to 2019 (t = 0 to 18 years) (in trillions of US$)

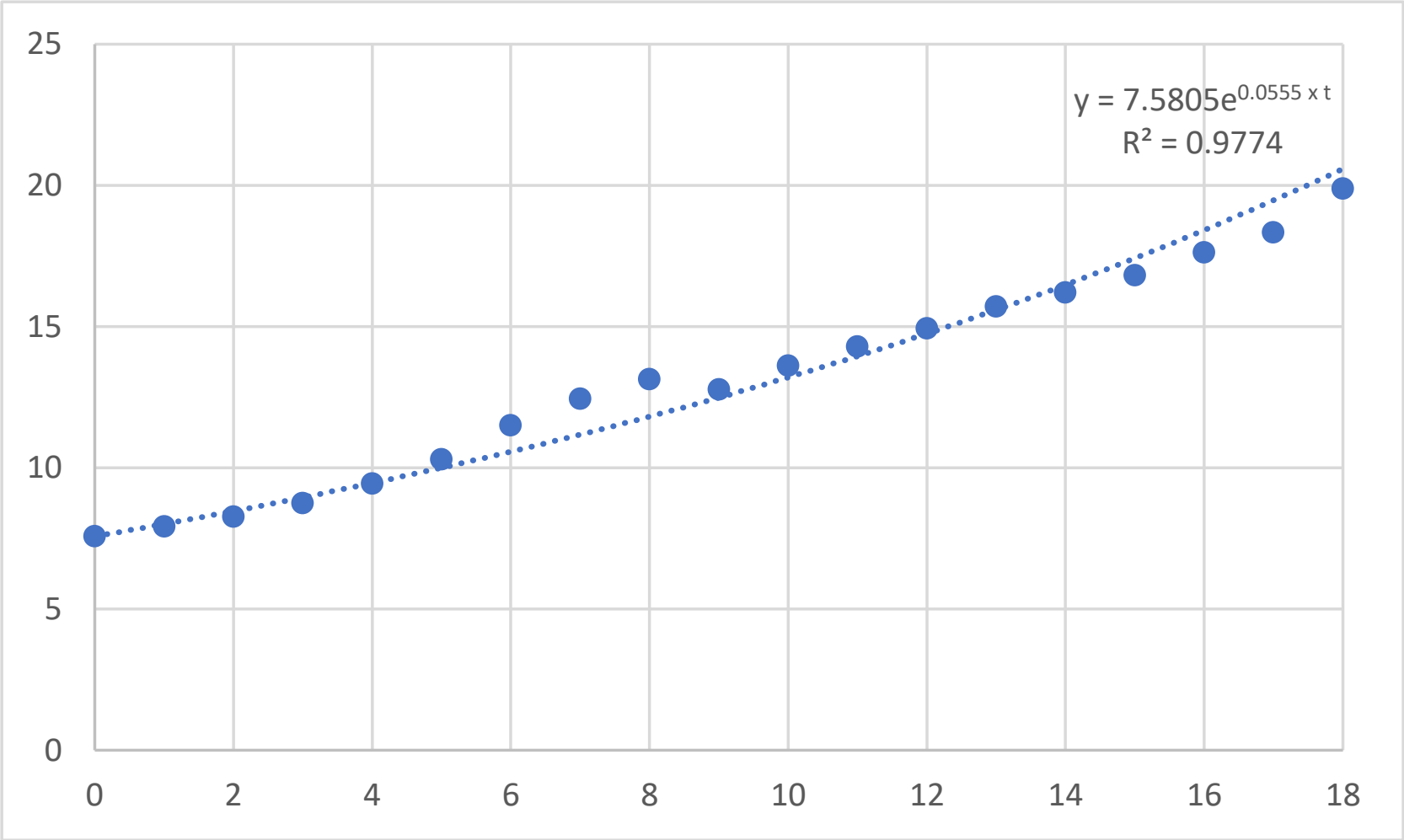

The information entropy of the US Broad Money Supply data for this time period is given by the equation:

$$Info\ Ent\ (US\ Broad\ Money\ Supply) = 0.0555\ x\ t$$

## 3.2 Information Entropy of the US GDP (2001 to 2019)

Using 2001 as the reference year (year = 0), the growth in US GDP for the period 2001 to 2019 (years 0 to 18) [3] is given in Figure 2.

Figure 2: The growth in US GDP for the period 2001 to 2019 (years 0 to 18)

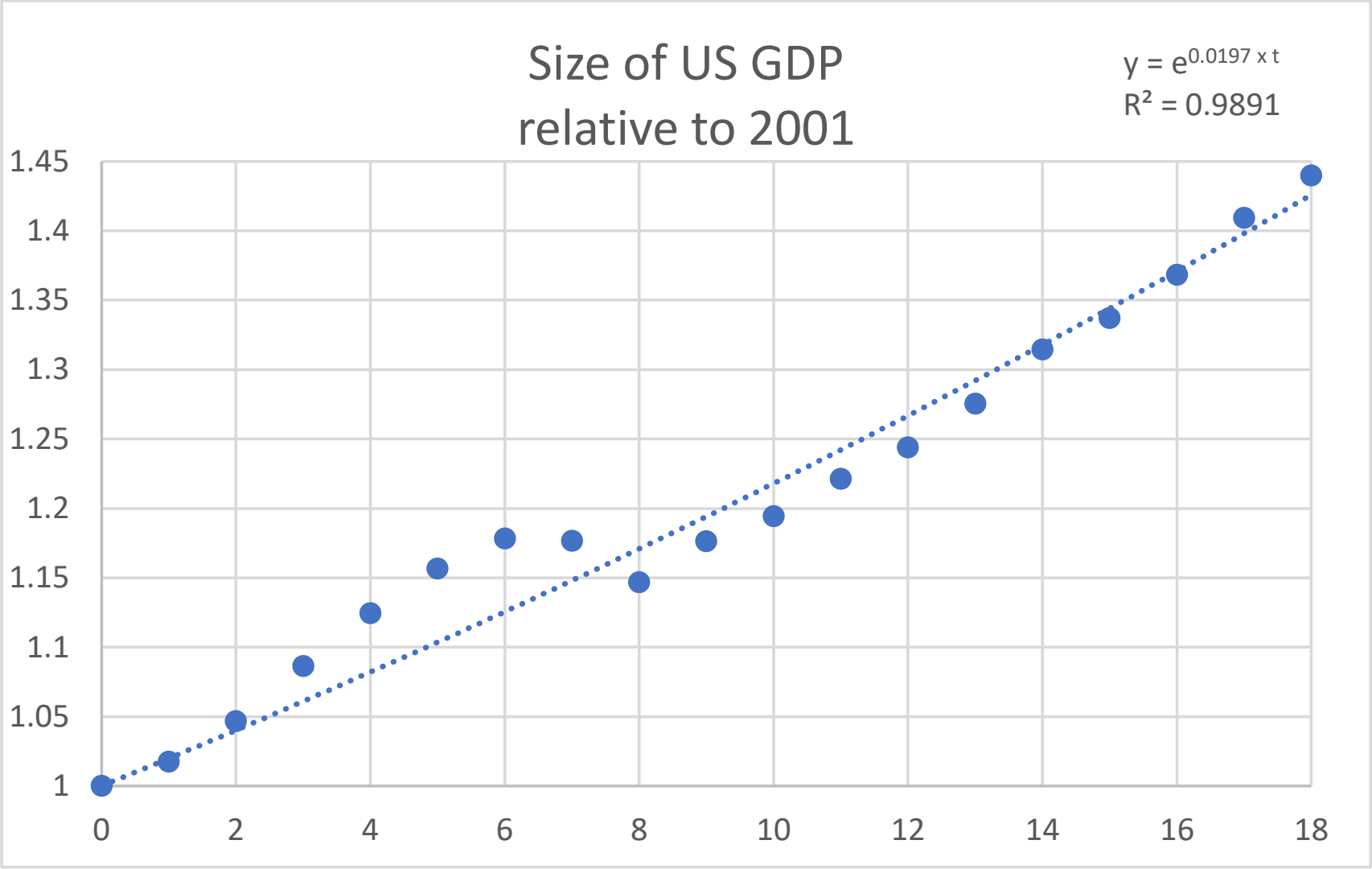

As can been seen, the US GDP over this period can be described by an exponential growth function, with a rate constant, λ = 0.0197. This corresponds to an average annual % increase in US GDP of 2.0% over this time period. The coefficient of determination ($R^2$) for the exponential characterization of the GDP data is approximately 99%.

The information entropy of the US GDP data for this time period is given by the equation:

$$Info\ Ent\ (US\ GDP) = 0.0197\ x\ t$$

The average annual rate of growth of US GDP (2.0%) was considerably lower than the average annual rate of growth of the US broad money supply (5.7%).

The information entropy of the US Broad Money Supply minus the information entropy of the US GDP for this time period is given by the equation:

$$Info\ Ent\ (US\ Broad\ Money\ Supply) - \ Info\ Ent\ (US\ GDP) = 0.0358\ x\ t$$

The information entropy of the US Broad Money Supply plus the information entropy of the US GDP for this time period is given by the equation:

$$Info\ Ent\ (US\ Broad\ Money\ Supply) + \ Info\ Ent\ (US\ GDP) = 0.0752\ x\ t$$

This is equivalent to a <u>combined</u> average annual growth rate of 7.8% (approx. 5.7% + 2.0%)

### 3.3 Exponential nature of US asset indices of interest

Using the US S&P 500 Index, as an example, its long-term exponential nature can be seen from its long-term time series, over the period 1928 to 2019 [4]. A semi-log plot for the entire time series is given in Figure 3.

Figure 3: The growth in the US S&P 500 Index for the period 1928 to 2019 (years 0 to 92) (semi-log plot)

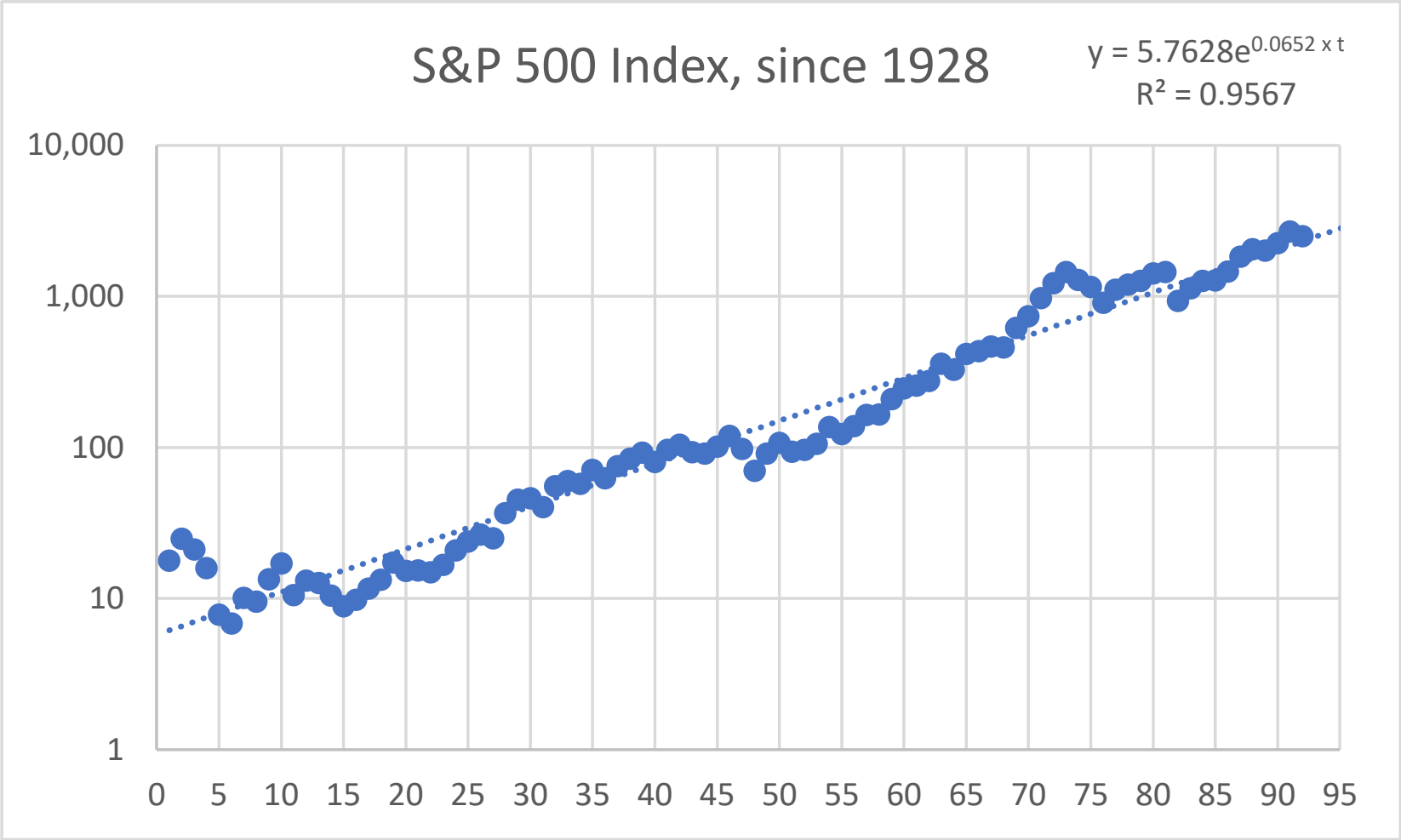

Applying an exponential growth function over the time period, provides a rate constant, λ = 0.0652. This corresponds to an average annual % increase in the S&P 500 Index of approx. 6.7% over this long time period. The coefficient of determination ($R^2$) for the exponential characterization of the S&P 500 Index data is approximately 96%.

## 3.4 Information Entropy of the US National Home Price Index (2001 to 2019)

Using 2001 as the reference year (year = 0), the growth in the US National Home Price Index for the period 2001 to 2019 (years 0 to 18) [5] is given in Figure 4.

Figure 4: The growth in the US National Home Price Index for the period 2001 to 2019 (years 0 to 18)

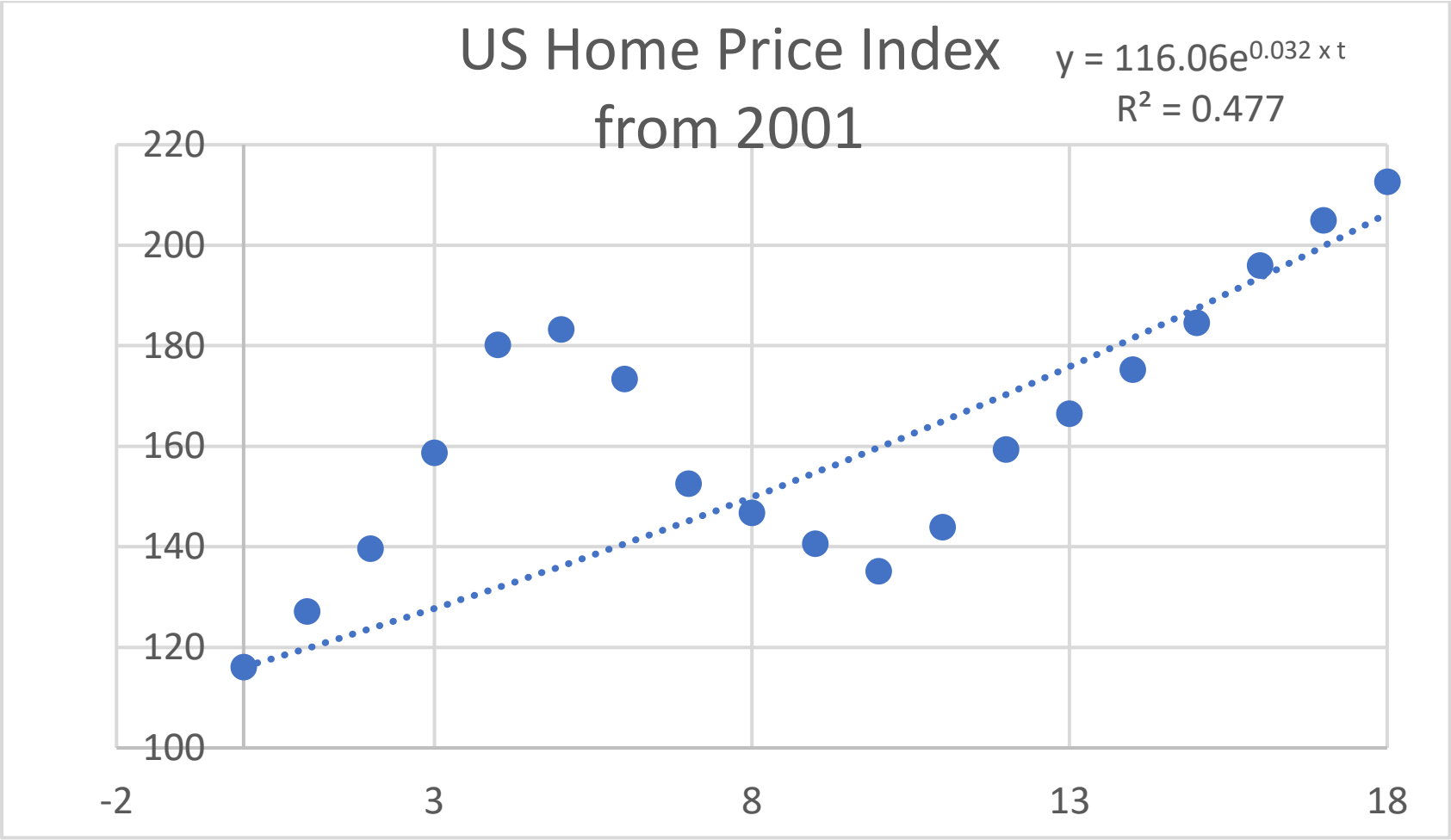

As can been seen, the US National Home Price Index was greatly impacted by the financial crisis of 2007-2008. Notwithstanding, applying an exponential growth function over the period, provides a rate constant, λ = 0.0320. This corresponds to an average annual % increase in the US National Home Price Index of 3.3% over this time period. The coefficient of determination ($R^2$) for the exponential characterization of the US National Home Price Index data is approximately 48%, reflecting the considerable decline in prices following the financial crisis.

The information entropy of the US National Home Price Index for this time period is given by the equation:

$$Info\ Ent\ (US\ Home\ Price\ index) = 0.0320\ x\ t$$

Notwithstanding the financial crisis of 2007-2008, this annual rate of increase in the US National Home Price Index over this time period is considerably greater than the corresponding growth in US GDP. However, it can be described as follow:

$$Info\ Ent\ (US\ Home\ Price\ index) =$$

$$Info\ Ent\ (US\ Broad\ Money\ Supply) - \ Info\ Ent\ (US\ GDP)\ -\ 0.0038 =\ 0.0320\ x\ t$$

This means that the US National Home Price Index over this time period had approx. 89% of the average annual rate of growth of the US Broad Money Supply minus the average annual rate of growth of US GDP.

## 3.5 Information Entropy of the US Russell 2000 Index (2001 to 2019)

Using 2001 as the reference year (year = 0), the growth in the US Russell 2000 Index for the period 2001 to 2019 (years 0 to 18) [6] is given in Figure 5.

Figure 5: The growth in the US Russell 2000 Index for the period 2001 to 2019 (years 0 to 18)

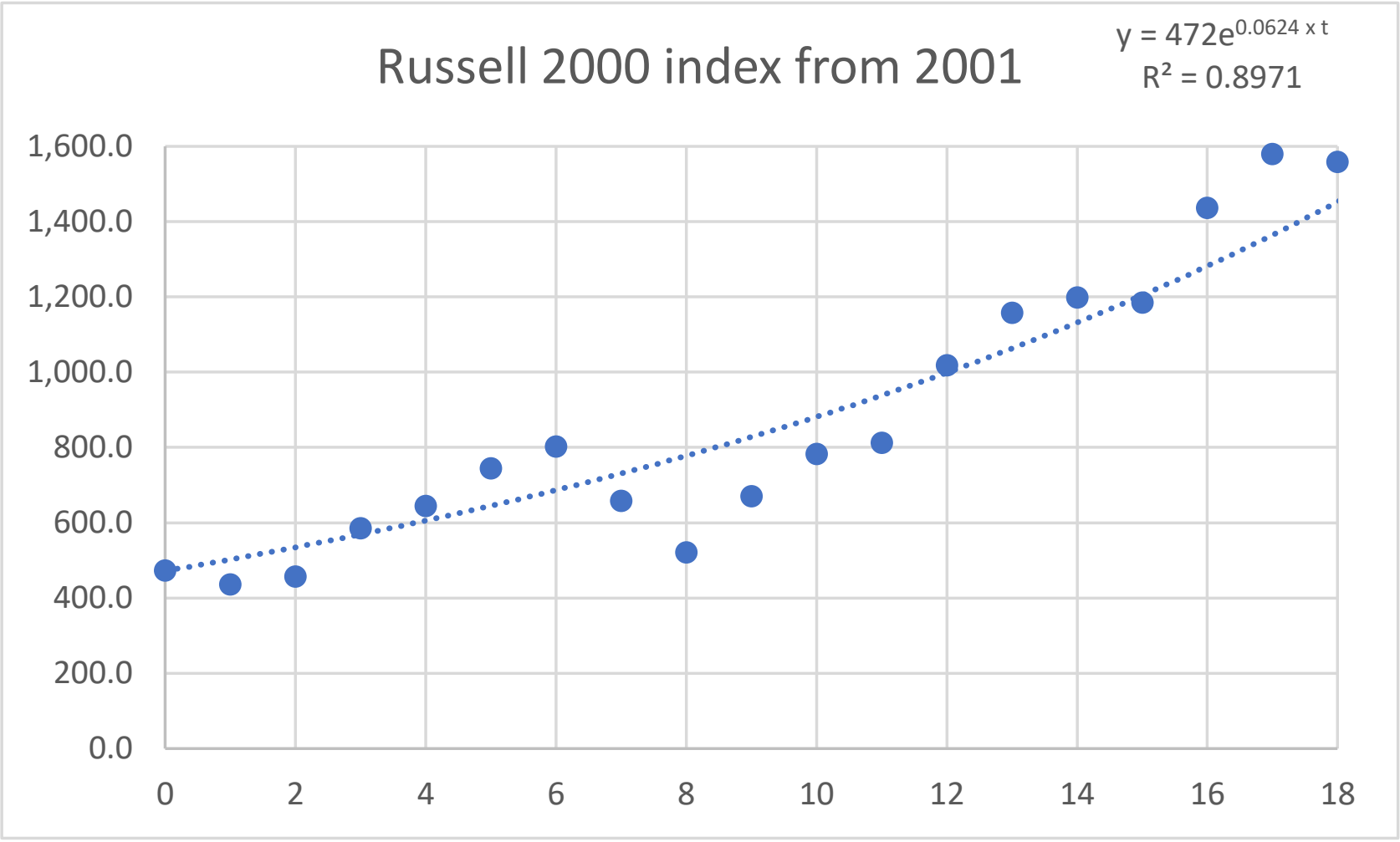

As can been seen, the US Russell 2000 Index was significantly impacted by the financial crisis of 2007-2008. Notwithstanding, applying an exponential growth function over the period, provides a rate constant, λ = 0.0624. This corresponds to an average annual % increase in the US National Home Price Index of 6.4% over this time period. The coefficient of determination ($R^2$) for the exponential characterization of the US Russell 2000 Index data is approximately 90%, reflecting the decline in prices following the financial crisis

The information entropy of the US Russell 2000 Index for this time period is given by the equation:

$$Info\ Ent\ (US\ Russell\ 2000\ index) = 0.0624\ x\ t$$

Notwithstanding the financial crisis of 2007-2008, this annual rate of increase in the US Russell 2000 Index over this time period is considerably greater than the corresponding growth in US GDP or the US Broad Money Supply. However, it can be described as follow:

$$Info\ Ent\ (US\ Russell\ 2000\ index) =$$

$$Info\ Ent\ (US\ Broad\ Money\ Supply) + Info\ Ent\ (US\ GDP) - 0.0128 = 0.0624\ x\ t$$

This means that the US Russell 2000 Index over this time period had approx. 83% of the average rate of growth of the US Broad Money Supply plus the average annual rate of growth of US GDP.

## 3.6 Information Entropy of the US S&P 500 Index (2001 to 2019)

Using 2001 as the reference year (year = 0), the growth in the US S&P 500 Index for the period 2001 to 2019 (years 0 to 18) [7] is given in Figure 6.

Figure 6: The growth in the US S&P 500 Index for the period 2001 to 2019 (years 0 to 18)

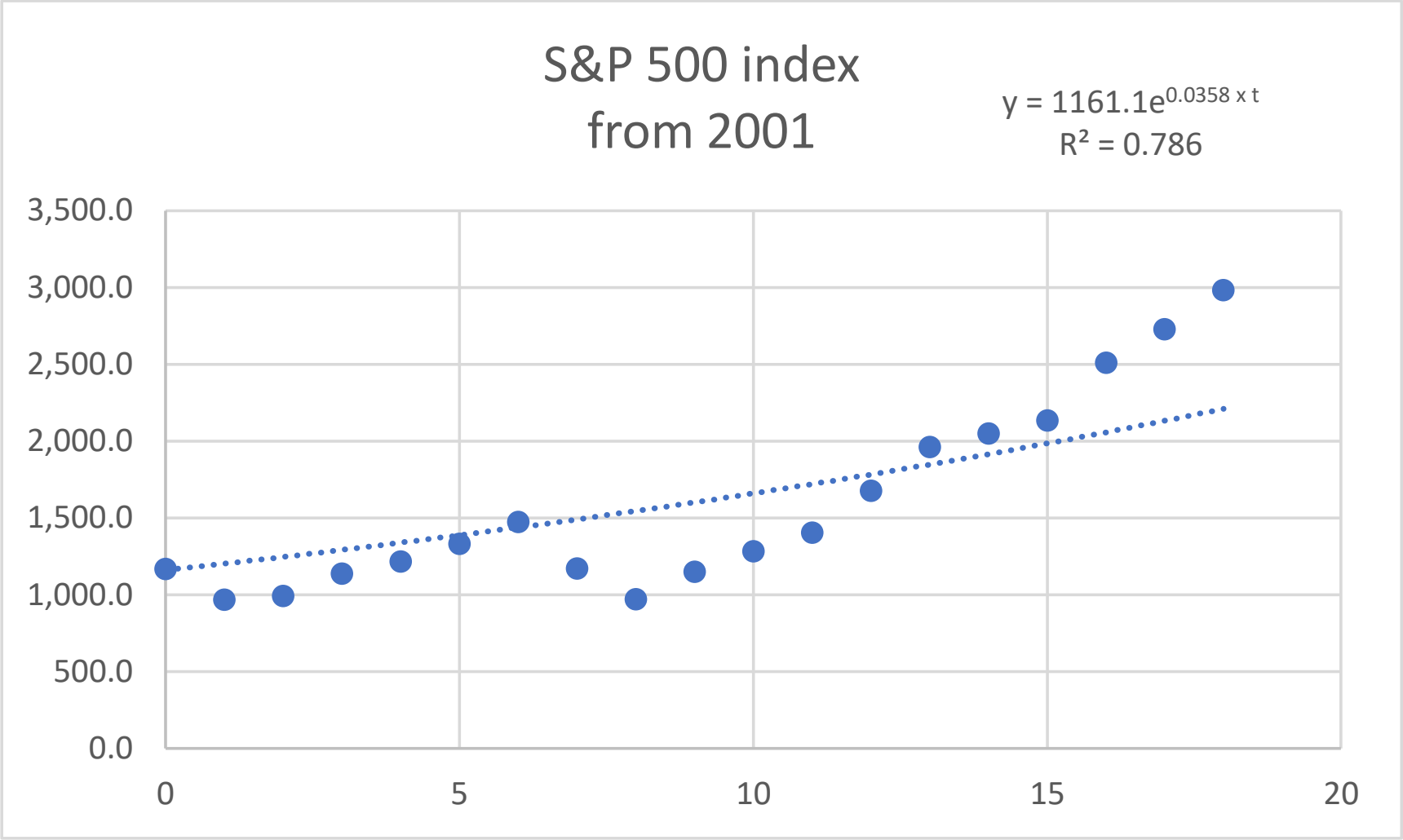

As can been seen, the US S&P 500 Index was greatly impacted by the financial crisis of 2007-2008. Notwithstanding, applying an exponential growth function over the period, provides a rate constant, λ = 0.0358. This corresponds to an average annual % increase in the US S&P 500 Index of 3.6% over this time period. The coefficient of determination ($R^2$) for the exponential characterization of the US National Home Price Index data is approximately 79%, in part, reflecting the decline in the index following the financial crisis.

The information entropy of the US S&P 500 Index for this time period is given by the equation:

$$Info\ Ent\ (US\ S\&P\ 500\ index) = 0.0358\ x\ t$$

Notwithstanding the financial crisis of 2007-2008, this annual rate of increase in the US S&P 500 Index over this time period is considerably greater than the corresponding growth in US GDP. However, it can be described as follow:

$$Info\ Ent\ (US\ S\&P\ 500\ index) =$$

$$Info\ Ent\ (US\ Broad\ Money\ Supply) - Info\ Ent\ (US\ GDP)\ \pm 0.0000 = 0.0358\ x\ t$$

This means that the US S&P 500 Index over this time period had 100% of the average annual rate of growth of the US Broad Money Supply minus the average annual rate of growth of US GDP.

## 3.7 Information Entropy of the NASDAQ Index (2001 to 2019)

Using 2001 as the reference year (year = 0), the growth in the NASDAQ Index for the period 2001 to 2019 (years 0 to 18) [8] is given in Figure 7.

Figure 7: The growth in the NASDAQ Index for the period 2001 to 2019 (years 0 to 18)

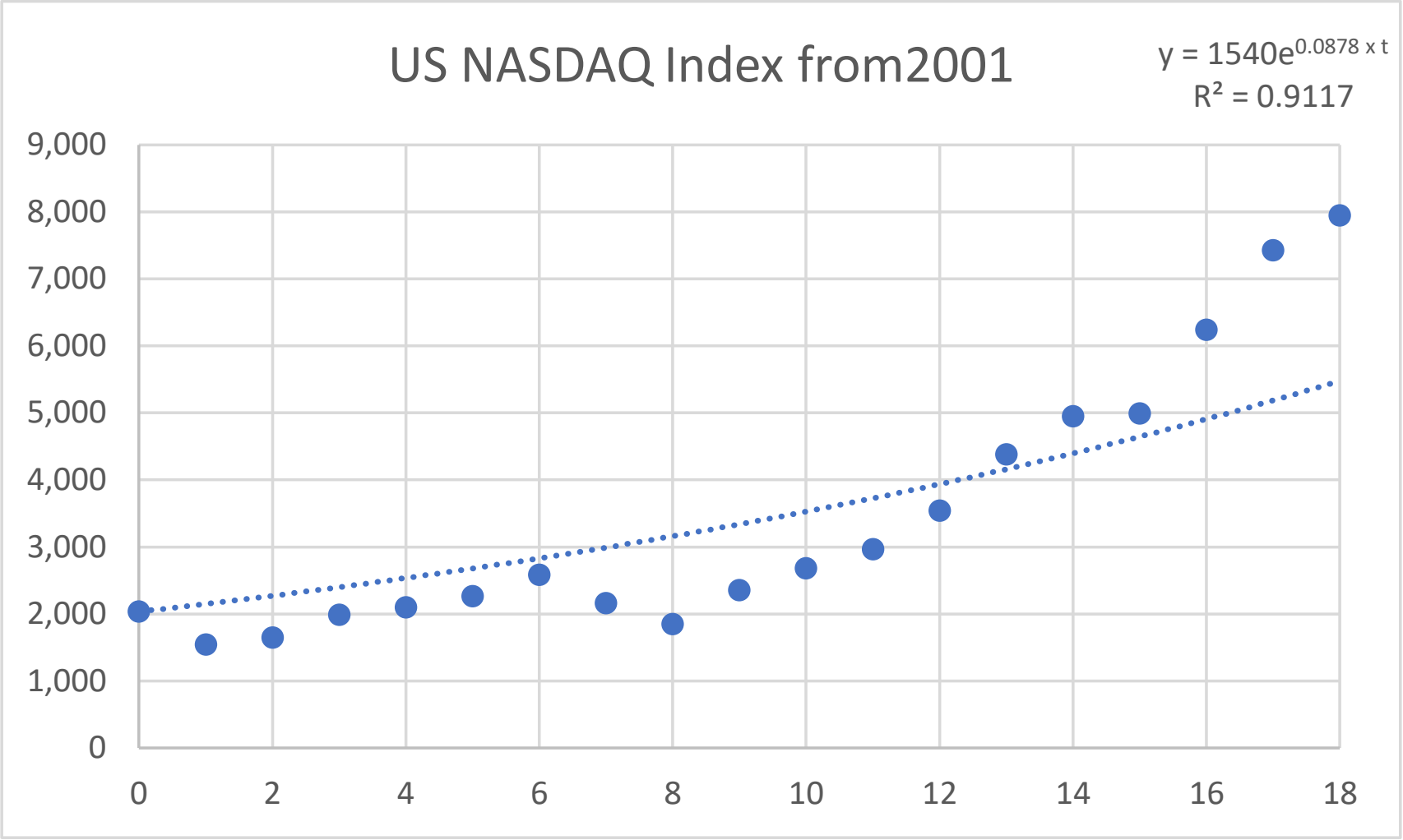

As can been seen, the NASDAQ Index was significantly negatively impacted by the financial crisis of 2007-2008. Notwithstanding, applying an exponential growth function over the period, provides a rate constant, λ = 0.0878. This corresponds to an average annual % increase in the NASDAQ Index of 9.2% over this time period. The coefficient of determination ($R^2$) for the exponential characterization of the NASDAQ Index data is approximately 91%, in part, reflecting the decline in the index following the financial crisis.

The information entropy of the NASDAQ Index data this time period is given by the equation:

$$Info\ Ent\ (NASDAQ\ index) = 0.0878\ x\ t$$

Notwithstanding the financial crisis of 2007-2008, this annual rate of increase in the NASDAQ Index over this time period is considerably greater than the corresponding growth in US GDP or the US Broad Money Supply. However, it can be described as follow:

$$Info\ Ent\ (NASDAQ\ index) =$$

$$Info\ Ent\ (US\ Broad\ Money\ Supply) + \ Info\ Ent\ (US\ GDP) + 0.0137 = 0.0878\ x\ t$$

This means that the NASDAQ Index over this time period had approx. 117% of the average annual rate of growth of the US Broad Money Supply <u>plus</u> the average annual rate of growth of US GDP.

## 3.8 Information Entropy of the US CPI (2001 to 2019)

The US Consumer Price Index (CPI) from 2001 to 2019 (time, 0 to 18 years) [9] is plotted in Figure 8. As can been seen, over this time period, the CPI can be well described using an exponential function. The coefficient of determination ($R^2$) of the fit is approx. 98% for the 19-year time series. Over this period the exponential rate-constant (λ) = 0.0203. This is equivalent to a mean annual rate of increase in the US CPI over this time period of 2.1%.

Figure 8: US CPI from 2001 to 2019 (years 0 to 18)

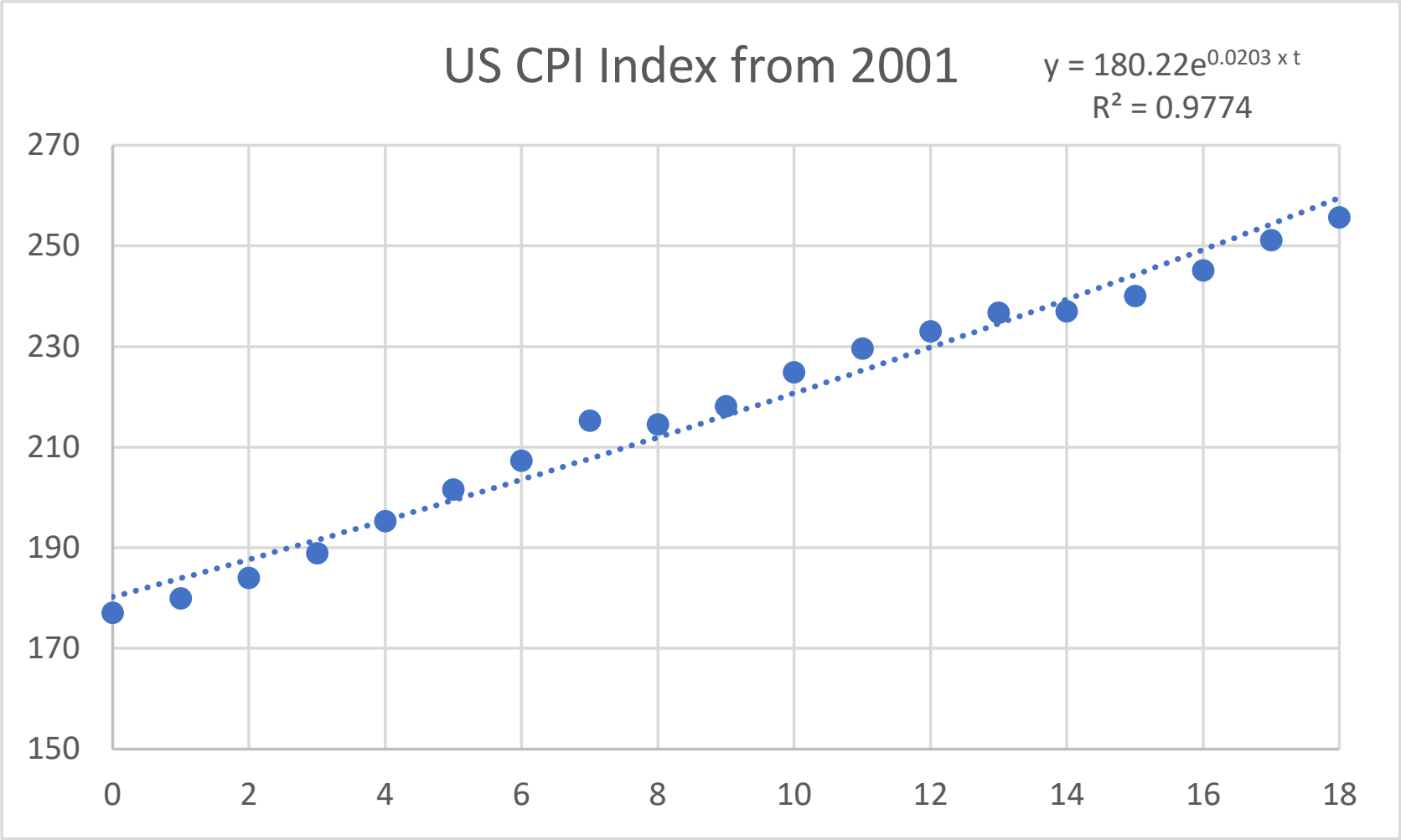

The information entropy of the US CPI over this time period is given by the equation:

$$Info\ Ent\ (US\ CPI) = 0.0203\ x\ t$$

## 4. Discussion

The information entropy methodology provides a rate-constant for the exponential growth and the associated information entropy for the time series under investigation. The information entropy can be considered to be a measure of the “velocity” of the time series. So, for the US GDP time series, it can be regarded as a measure of the “velocity” of US GDP, over the period 2001 to 2019.

The period 2001 to 2019 was purposefully selected as it contained within it the financial crisis of 2007-2008. This crisis had a significant negative impact on the US asset indices. However, notwithstanding the financial crisis of 2007-2008, US real GDP increased exponentially over the period 2001 to 2019, with an average annual growth rate of approximately 2%. In addition, inflation was held steady at an average annual growth rate of approximately 2%. However, over this time, the average annual rate of growth of US GDP was considerably lower than the average annual rate of growth of the US broad money supply (5.7%).

Over this time period, the time series for the major asset indices grew, notwithstanding the negative impact of the financial crisis of 2007-2008. The primary objective of this paper was to investigate whether the growth in the major US asset indices could be a function of the US broad money supply and/or US GDP.

The information entropies of the four different US asset indices over the time period 2001 to 2019 are compared in Figure 9.

Figure 9: Comparison of the information entropies of the four* different US indices over the period 2001 to 2019, with 2001, as time = 0, with that of the US CPI

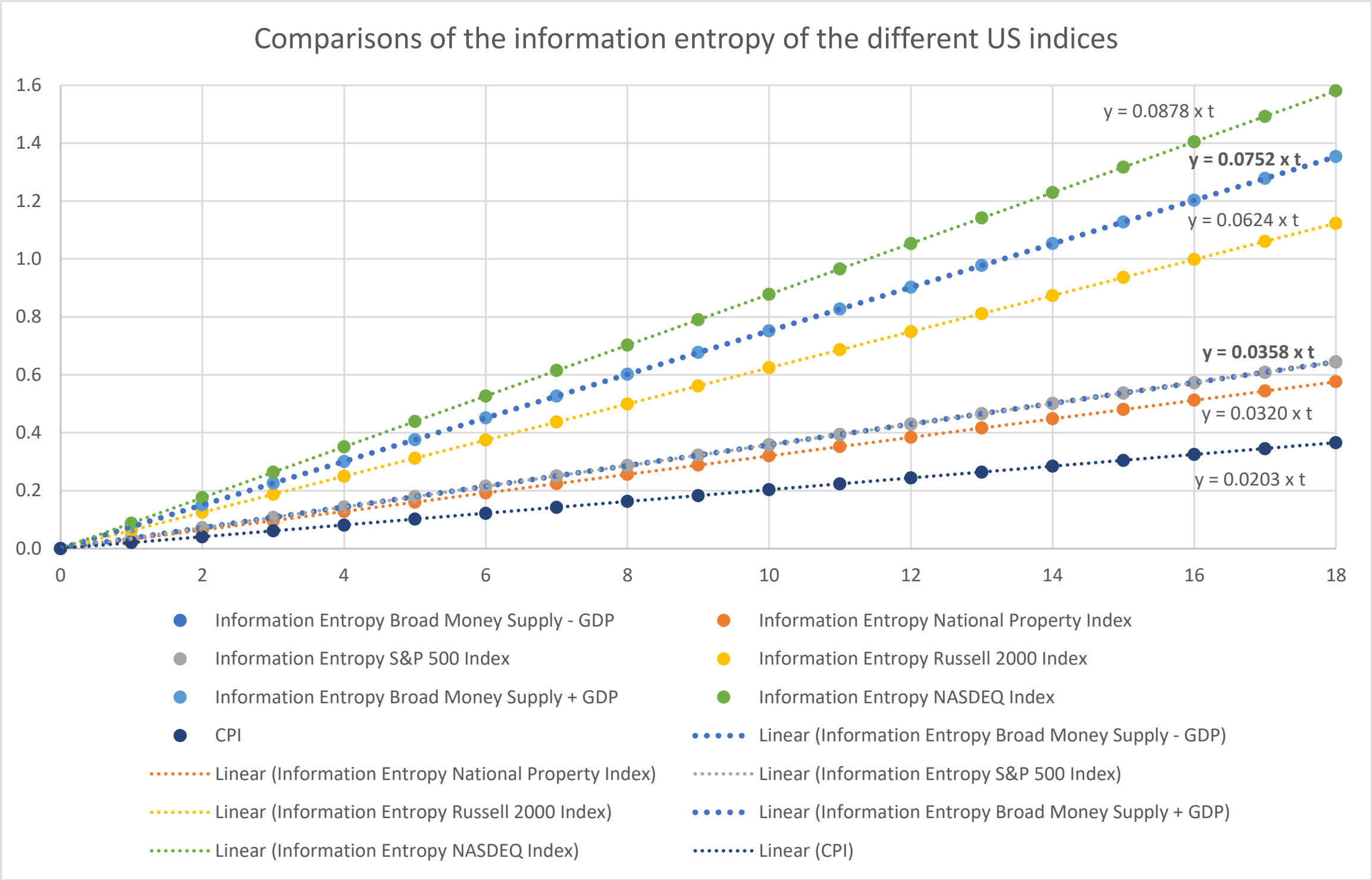

*Note: there is a complete overlap of the Info Ent for S&P 500 index with that for the US Broad Money Supply – US GDP

It can be seen that they appear to fall into two separate groups:

(1) Those assets that would appear to be function of the US Money Supply multiplied by US GDP, given by the equation:

$$Info\ Ent\ (US\ asset\ index) = Info\ Ent\ (\ US\ Broad\ Money\ Supply) + Info\ Ent\ (US\ GDP) \pm constant$$

These assets comprise the Russell 2000 index and the NASDAQ index.

(2) Those assets that would appear to be function of the US Broad Money Supply divided by US GDP given by the equation:

$$Info\ Ent\ (US\ asset\ index) = Info\ Ent\ (\ US\ Broad\ Money\ Supply) - \ Info\ Ent\ (US\ GDP) \pm constant$$

These assets comprise the US National Property index and the S&P 500 index.

However, what all four indices share in common is that, over the period 2001 to 2019, they would appear to be a function of the US Broad Money Supply. This would seem plausible for the stock market indices, since an expanding US Broad Money Supply, provides the potential for increasing sums of monies to be invested in these indices, particularly by large investment institutions. However, a simple correlation relationship cannot be ruled out.

What was not the *a priori* expectation was the finding that the average rate of growth of US National Property index would be considerably greater than that of US GDP, over the time period that included the financial crisis of 2007-2008. In addition, real US annual average wages and salaries increased very modestly over this time period [10].

What was the *a priori* expectation was that average annual rate of increase of the Russell 2000 index would be a function of the average annual rate of growth of US Broad Money Supply plus the average annual rate of growth of US GDP, over the period 2001-2019. This was because that in addition to the growth in the US broad money supply over this period, the profits of the US companies that comprise the Russell 2000 index would be expected to increase, as the US GDP increased exponentially over this time period.

What might be useful follow-up research would be to investigate the time series for each of the four US asset indices over the period 1981 to 2000, using the information entropy methodology, as this would be a time period that did not contain a financial crisis. In addition, there may be further scope to apply this information entropy methodology to investigate other topics in the field of macro-economics.

The approach taken in this paper might appear to be loosely based on the concept of the "velocity of money" [11], in which:

$$"velocity\ of\ money" = GDP/Money\ Supply$$

This equation can be rearranged as follow:

$$log_e\ (\text{Money Supply}) - log_e\ (\text{GDP}) = -\ log_e\ (\text{"velocity of money"})$$

However, any similarities between the two approaches are merely superficial.

This paper has shown that the information entropy approach can be applied to the analysis of macroeconomic time series data to address research questions. Other existing methodologies could also be used to address the research questions investigated in this paper.

The information entropy approach is very straight forward to implement and it can convert relationships under investigation into a "linearized" form. For example,

$$\begin{aligned} Info\ Ent\ (US\ stock\ market\ index) \\ &= Info\ Ent\ (US\ Broad\ Money\ Supply)\ \pm Info\ Ent\ (US\ GDP) \\ &\pm constant \end{aligned}$$

is equivalent to the combined set of four equations given below:

(1) $US\ stock\ market\ index =\ US\ Broad\ Money\ Supply\ x\ US\ GDP\ x\ constant$

(2) $US\ stock\ market\ index =\ US\ Broad\ Money\ Supply\ x\ US\ GDP\ /\ constant$

(3) $US\ stock\ market\ index = (US\ Broad\ Money\ Supply\ /\ US\ GDP)\ x\ constant$

(4) $US\ stock\ market\ index = (US\ Broad\ Money\ Supply\ /\ US\ GDP)\ /\ constant$

The information entropy approach evaluates the dynamic nature of a system. The information entropy yields the "velocity" of the process. It's derivative with time would yield the "acceleration" (or "de-acceleration") of the process. In the current paper, the "acceleration" of the US broad money supply, US GDP and US RPI could be regarded to be approximately linear over the time period of investigation. This may not be expected to be the case with other macroeconomic time series or over other time periods. For these reasons, the information entropy approach may be a useful additional methodology to those other existing data analysis techniques currently used in macroeconomics.

## 5. Conclusion

Over the period 2001 to 2019, the average annual rate of growth of US GDP (2.0%) was considerably lower than the average annual rate of growth of the US broad money supply (5.7%). Over this time period, the main determinant of the average growth rate for all four US asset indices would appear to be the growth in the US broad money supply. The rate of growth in the US Russell 2000 stock index (an index of smaller US companies) and the NASDAQ index would appear to be a function of the combined positive effects of both the growth rate in the US Broad Money Supply and the growth rate of US GDP. The approach taken in this paper may appear to be very loosely based on the concept of the "velocity of money". However, any similarities between the two approaches are merely superficial.

## Supplementary materials

There are no supplementary materials.

## Acknowledgements

The author gratefully acknowledges those academic and non-academic people who kindly provided feedback on earlier drafts of this paper. No financial support was received for any aspect of this research.

## References

[1] Lacey, LF. Characterization of the probability and information entropy of a statistical process with an exponentially increasing sample space. arXiv:2105.14193. Accessed on 01 June 2021 from arxiv.org. url: https://arxiv.org/abs/2105.14193

[2] "The broad money supply of US$ from 2001 to 2019". Monetary, Broad Money, US Dollars, International Financial Statistics (IFS). Accessed on 13 April 2021 from International Monetary Fund (IMF). url: https://data.imf.org/?sk=B83F71E8-61E3-4CF1-8CF3-6D7FE04D0930&sId=1393552803658

[3] "Gross Domestic Product for United States, Percent Change, Annual, Not Seasonally Adjusted". Accessed on 15 April 2021 from Federal Reserve Bank of St. Louis. url: https://fred.stlouisfed.org/series/USANGDPRPCH

[4] "The S&P 500 (1928 to 2019)". Accessed on 18 April 2021 from TheStreet.com. url: https://www.thestreet.com/investing/annual-sp-500-returns-in-history

[5] "S&P/Case Shiller U.S. National Home Price Index from 2000 to 2019". Accessed on 15 April 2021 from Statista.com. url: https://www.statista.com/statistics/199360/case-shiller-national-home-price-index-for-the-us-since-2000/

[6] "US Russell 2000 Index from 2000 to 2019". Accessed on 15 April 2021 from MarketWatch. url: https://www.marketwatch.com/investing/index/rut/download-data?startDate=1/3/2000&endDate=12/31/2020

[7] "US S&P 500 Index from 2001 to 2019". Accessed on 15 April 2021 from Yahoo Finance. url: https://finance.yahoo.com/quote/%5EGSPC/history?period1=980640000&period2=1577491200&interval=1mo&filter=history&frequency=1mo&includeAdjustedClose=true

[8] "NASDAQ Index from 2001 to 2019". Accessed on 15 April 2021 from Yahoo Finance. url: https://finance.yahoo.com/quote/%5EGSPC/history?period1=980640000&period2=1577491200&interval=1mo&filter=history&frequency=1mo&includeAdjustedClose=true

[9] "US Consumer Price Index (CPI) From 1903". Federal Reserve Bank of Minneapolis. Accessed

on 15 April 2021 from url: https://www.minneapolisfed.org/about-us/monetary-policy/inflation-calculator/consumer-price-index-1913-

[10] Desilver D. “For most U.S. workers, real wages have barely budged in decades”. Pew Research Centre, August 7 (2018). Accessed on 15 April 2021 from url: https://www.pewresearch.org/fact-tank/2018/08/07/for-most-us-workers-real-wages-have-barely-budged-for-decades/

[11] “Velocity of Money”. Accessed on 13 April 2021 from Investopedia. url: https://www.investopedia.com/terms/v/velocity.asp